\documentclass[tradiabstract]{aa}

\usepackage{natbib}
\usepackage{graphicx}
\usepackage{txfonts}
\usepackage{epsfig}
\usepackage{longtable}
\usepackage{lscape}
\usepackage{amssymb}
\usepackage{soul}
\usepackage[colorlinks=true,citecolor=blue]{hyperref}
\usepackage{refcount}
\usepackage{txfonts}
\usepackage{epsfig}
\usepackage{longtable}
\usepackage{lscape}
\usepackage{amssymb} 
\usepackage[usenames,dvipsnames,svgnames,table]{xcolor}
\usepackage[colorlinks=true,citecolor=blue]{hyperref} 
\usepackage{refcount}
\usepackage{lipsum}
\usepackage{subcaption}         
\usepackage{placeins}           

\def \mysrc {4U~1909+07}

\urldef\xrtpos\url{https://www.swift.ac.uk/user_objects/docs.php#posform}

 \definecolor{arancio}{rgb}{1,0.5,0}
 \definecolor{viola}{rgb}{0.7,0,1}
 \definecolor{verde}{rgb}{0.2,0.7,0.7}
\definecolor{cobalt}{rgb}{0.0, 0.28, 0.67}
\definecolor{airforceblue}{rgb}{0.36, 0.54, 0.66}
\definecolor{ballblue}{rgb}{0.13, 0.67, 0.8}
\definecolor{battleshipgrey}{rgb}{0.52, 0.52, 0.51}
\definecolor{darkgreen}{rgb}{0.0, 0.2, 0.13}

\def \rxte {{\em RXTE}}

\def \suzaku {{\em Suzaku}}
\def \sw {{\em Swift}}
\def \nustar {{\em NuSTAR}}

\def \hcm {\hbox {\ifmmode $ atom cm$^{-2}\else atom cm$^{-2}$\fi}}

\def \apj {ApJ}

\def \natas {Nature Astronomy}
\def \mnras {MNRAS}

\def \ssr {SSRv}

\begin{document} 

\title{\emph{Swift}/XRT monitoring of the orbital and superorbital modulations in 4U~1909+07 }

\author{P.~Romano\inst{\ref{oab}}              
    \and
H.I.~Cohen \inst{\ref{cua},\ref{uom},\ref{gsfc}}   
    \and
E.~Bozzo \inst{\ref{geneve},\ref{oab}}        
    \and
N.~Islam \inst{\ref{uom},\ref{gsfc}}              
    \and
A.~Lange   \inst{\ref{gwu},\ref{uom},\ref{gsfc}}     
    \and
R.H.D.~Corbet \inst{\ref{cresst},\ref{gsfc},\ref{mic}}   
    \and   
J.B.~Coley  \inst{\ref{how},\ref{apl}}    
    \and  
K.~Pottschmidt \inst{\ref{uma},\ref{apl}} 
 }  

\institute{
    INAF -- Osservatorio Astronomico di Brera, 
                  Via E.\ Bianchi 46, I-23807, Merate, Italy
                  \email{patrizia.romano@inaf.it}\label{oab}
            \and
   Catholic University of America, 620 Michigan Ave NE, Washington, DC 20064, USA \label{cua} 
              \and 
    Center for Space Science and Technology, University of Maryland, Baltimore County, 
    1000 Hilltop Circle, Baltimore, MD 21250, USA \label{uom}   
              \and 
     X-ray Astrophysics Laboratory, NASA Goddard Space Flight Center, Greenbelt, 
     MD 20771, USA \label{gsfc}
            \and
    Department of Astronomy, Universit\'e de Gen\`eve, 16 chemin d'\'Ecogia, 1290 Versoix, 
    Switzerland \label{geneve} 
            \and    
    George Washington University, Washington, DC 20052, USA \label{gwu}   
              \and 
    CRESST and CSST, University of Maryland, Baltimore County, 1000 Hilltop Circle, Baltimore, 
    MD 21250, USA \label{cresst} 
            \and
    Maryland Institute College of Art, 1300 W Mt Royal Ave, Baltimore, 
    MD 21217, USA \label{mic}
    \and
    Department of Physics and Astronomy, Howard University, Washington, DC 20059, USA \label{how}
\and   CRESST/Astroparticle Physics Laboratory, Code 661, NASA Goddard Space Flight Center, Greenbelt Road, MD 20771, USA \label{apl}
    \and
     University of Maryland, Baltimore County, MD 21250, USA \label{uma}
     }

\date{Received 18 March 2025 /  Accepted  2 May 2025} 

\abstract{We report on an observational campaign performed with \sw/XRT on the wind-fed supergiant X-ray binary 4U~1909+07 to investigate the nature of the orbital and superorbital modulation of its X-ray emission. A total of 137 XRT observations have been carried out, summing up to a total effective exposure time of 114~ks and covering a total of 66 orbital and 19 superorbital cycles of the source. The XRT data folded on the orbital period of the source confirmed and improved the previously reported variability in intensity and absorption column density, which can be ascribed to the neutron star accreting from the wind of its B supergiant companion across a fairly circular orbit. The XRT data folded on the superorbital period did not provide evidence of significant variations in either the absorption column density and/or the power-law photon index. This may be due to a significant weakening of the superorbital modulation during the times when the XRT observations were carried out, as confirmed by the BAT dynamic power spectrum.
We discuss the implications of these findings within the corotating interaction region model proposed to interpret the superorbital variability in wind-fed supergiant X-ray binaries.}


\maketitle

 \section{Introduction\label{supero2:intro}}

\vfill

Superorbital modulation has been detected in a variety of X-ray binaries hosting both black hole and neutron star accretors. These modulations manifest themselves as periodic changes in the X-ray luminosity of the system over timescales that are typically a few times longer than those associated with the orbital revolutions \citep[see, e.g.,][and references therein]{CorbetKrimm2013:superorbital,Corbet2021:superorbital}. In disk-fed binaries, the superorbital modulations are often ascribed to the precession of the accretion disk \citep[see, e.g., the case of Her\,X-1, SMC X-1, and LMC X-4;][and references therein]{Heyl2024:herx1}. In wind-fed supergiant X-ray binaries, where the primary star is in most cases an OB supergiant, interpretations of these phenomena are complicated because of the lack of sufficiently large and long-lasting (months to years) structures that could lead to (quasi-) periodic modulations of the mass accretion rate. 

\citet{Bozzo2017:superorbital} proposed that superorbital modulations in supergiant wind-fed binaries could be ascribed to the presence of the so-called Corotating Interaction Regions (`CIRs'), which are large-scale structures protruding from the photosphere of the massive star and extending out to several stellar radii. These structures are known to result in long-lasting ($\sim$years) density and velocity variations in the surroundings of the OB supergiant and could lead to significant modulation of the mass accretion rate onto a compact object with a periodicity that can easily exceed by a few times the binary orbital period. The model predicts that the encounter between the CIR and the compact object leads to a higher mass accretion rate combined with an enhanced absorption measured along the line of sight to the binary. This occurs because the CIR typically has a density that is higher by up to a factor of few times than the surrounding stellar wind, and a velocity which could be different from that of the smooth stellar wind  \citep[see, e.g.,][and references therein]{Lobel2008}. Depending on different possible geometrical configurations, the model by \citet{Bozzo2017:superorbital} shows that an enhanced absorption column density could be measured before or after the increased X-ray luminosity produced by the compact object interacting with the higher density of the CIR. The augmented X-ray luminosity could also result in subsequent short-term decreases in the local absorption column density, if a significant photoionisation of the CIR material occurs during the encounter with the compact object. 

In order to test the applicability of the CIR model to wind-fed supergiant X-ray binaries, \citet{Romano2024:superorbital} initiated monitoring observations of several of these system with the X--ray Telescope \citep[XRT,][]{Burrows2005:XRT}, the X-ray narrow field instrument on-board the the Neil Gehrels \sw\ Observatory \citep[][]{Gehrels2004}. The combination of high sensitivity, soft X-ray energy coverage, and scheduling flexibility of XRT proved very effective in probing the physics of superorbital modulations. XRT observations, as short as few hundreds of seconds, can be periodically scheduled to probe the properties (intensity and spectral energy distribution) of the source X-ray emission at different orbital and superorbital phases over many modulation cycles. This allows us to average out the short-term variability \citep[on hundreds to thousands seconds due to the clumpy wind accretion; see, e.g.,][and references therein]{Martinez-nunez2017,Kretschmar2019} and unveil the changes in the emission properties strictly related to the orbital and superorbital variations.  

In this paper, we extend our previous work \citep{Romano2024:superorbital} by presenting a ten-month monitoring campaign performed in 2024 with  \sw/XRT on the supergiant wind-fed X-ray binary \mysrc.\ We report on the search for flux and spectral variability in the X-ray emission from the source along both the orbital and superorbital cycles. We also exploit the rich archival data provided by the Burst Alert Telescope \citep[BAT,][]{Barthelmy2005:BAT} Transient Monitor \citep[][]{Krimm2013:BATTM} to update previously reported results on the source, performing a long term investigation of the presence and strength of its orbital and super-orbital modulation at hard X-rays across the past $\sim$20~years.

     \section{4U~1909+07} \label{supero2:4u}
 
\mysrc\ is a supergiant wind-fed high mass X-ray binary hosting a B0-3 I supergiant star \citep[][]{Martinez-nunez2015A:4U1909} and a neutron star accreting from its stellar wind. The system has an orbital period of $P_{\rm orb}=4.400\pm0.001$\,d.  Pulsations from the neutron star could be identified at $P_{\rm spin} \sim 605$\,s already in the early 2000's \citep[][]{Wen2000:1909,Levine2004:1909}, and were later demonstrated to show long-term changes \citep[][see also Coley et al., in prep., for the most recently evaluated trend]{Furst2011:4U1909,Furst2012:4U1909,Jaisawal2013:4U1909,Jaisawal2020:4U1909,Islam2023_superorbital}. \citet[][]{Jaisawal2013:4U1909} reported a possible absorption-like feature at 44\,keV in \suzaku\ observations identified as cyclotron resonant scattering feature (CRSF) and suggesting a neutron star surface magnetic field of $3.8\times10^{12}$\,G. The presence of this tentative CRSF, however, was not confirmed by subsequent \nustar\ and AstroSat observations \citep[][]{Jaisawal2020:4U1909,Islam2023_superorbital}. A study of the system X-ray luminosity variations as a function of the orbital phase was first reported by \citet[][]{Levine2004:1909} using \rxte/PCA  data. 

The discovery of a superorbital modulation at $\sim$15.2\,d was made possible by the exploitation of long-term \sw/BAT data and reported for the first time by \citet{CorbetKrimm2013:superorbital}. More recently, \citet[][]{Islam2023_superorbital} combined long term monitoring of the superorbital modulations using the BAT data, and pointed observations \sw\ and \nustar\ data, to probe possible changes in the intensity of the superorbital variability over time, as well as the broad-band spectral changes that might occur at different superorbital phases. They found that the intensity of the superorbital modulation in the source varies significantly over time and there are periods were it can be barely detected in the BAT data. They also found marginal evidence of a possible correlation between the changes in the spin-periods of 4U~1909+07 with the strength of the superorbital modulations, similar to what has been reported for 4U~1538$-$52 \citep[][]{Corbet2021:superorbital}. 
However, it was concluded that the pulse-period measurements in 4U~1909+07 are still too sparse to draw a firm conclusion. 
The few pointed \sw\ and \nustar\ data collected at one of the expected maximum and minimum of the superorbital variability did not reveal significant spectral variations in either the power-law slope or the recorded absorption column density
(see also Coley et al., in prep.).

  \section{Data reduction and analysis} \label{supero2:dataredu}
 
         \subsection{\emph{Swift}/XRT}  \label{supero2:data_xrt}

For this work, we combined two sets of {\it Swift}/XRT observations. The first set was collected as part of the \emph{Swift} GI program (Target ID 97732, PI N.\ Islam) as monitoring of 1\,ks observations twice per week from 2024-05-27 to 2024-10-10, for a total of 39 observations and an exposure of 33\,ks. 
The second set is a long-term monitoring campaign (Target ID 16436, PI: P.\ Romano) consisting of 1\,ks observations performed 3 times a week from 2024-02-18 to 2024-12-04 (MJD\,60358--60648), for a total of 98 observations and $\sim 83$\,ks.  We only considered data collected in photon counting (PC) mode,  yielding 137 observations for a total effective exposure time on the target of 114\,ks. 

We adopt P$_{\rm orb}$=4.4006$\pm$0.0006\,d for the orbital period     \citep[][with the phase zero being the minimum of the folded light-curve flux, at MJD\,50439.919]{Wen2000:1909,Levine2004:1909}, and P$_{\rm sup}$=15.196$\pm$0.004\,d for the superorbital period \citep[][with phase zero being the time of the maximum of the folded light-curve, at MJD\,59502.1]{Islam2023_superorbital}. Our observations, therefore, as reported in Table~\ref{supero2:tab:swift_xrt_log_u1909},  cover $\sim66$ orbital, and $\sim19$ superorbital cycles. This is graphically shown in Fig.~\ref{supero2:fig:checks_1909}, using visualization techniques already exploited in our previous paper on superorbital variabilities with XRT \citep{Romano2024:superorbital}.

\setcounter{figure}{0}
\begin{figure}
 \vspace{-0.4truecm}

 \hspace{-0.9truecm}
    \includegraphics[width=10cm,angle=0]{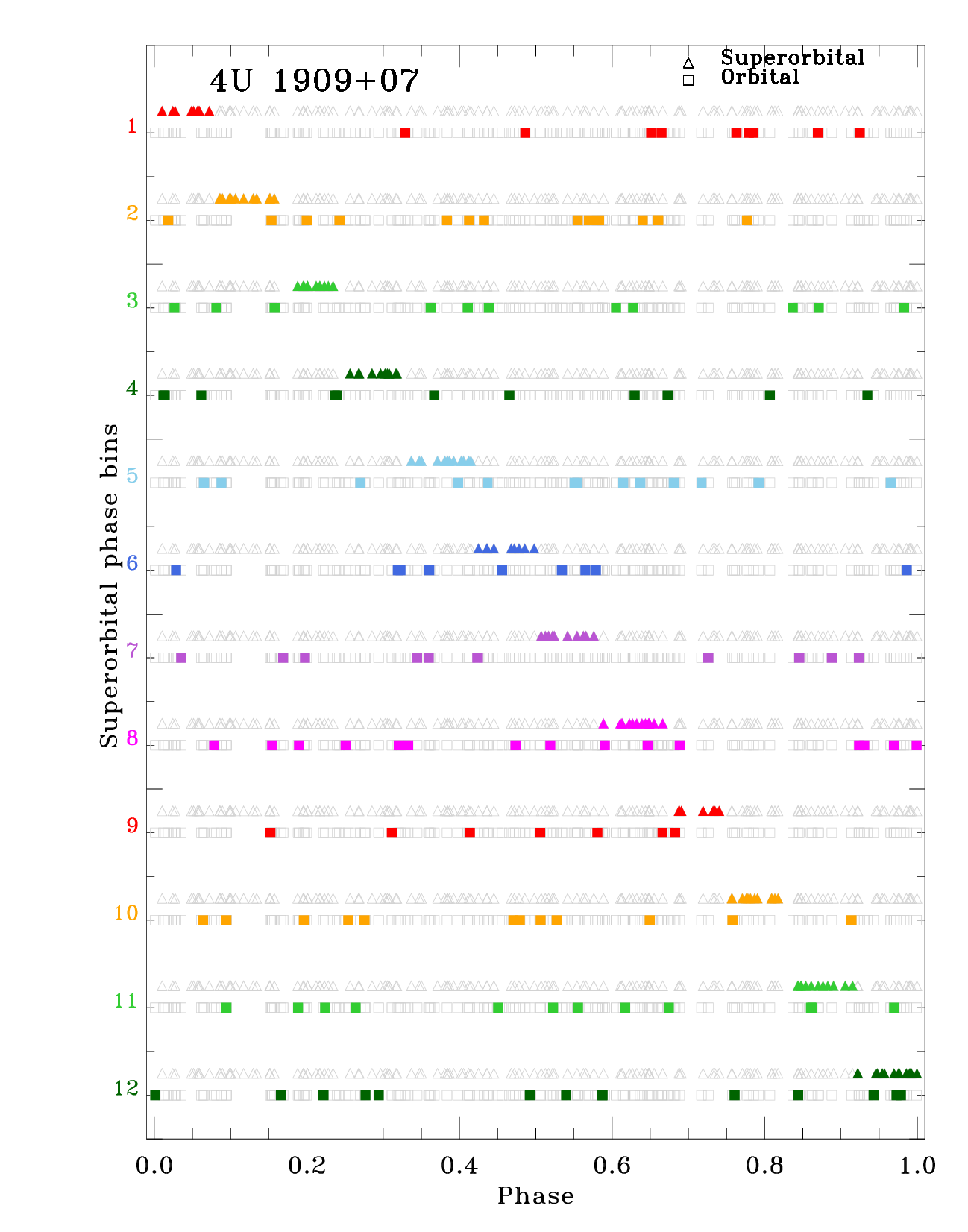}
  \caption{Distribution of the observations collected on \mysrc\ (on the x-axis) in  superorbital phase (grey empty upward-pointing triangles) and  orbital phase (grey empty squares)  as a function of superorbital phase bin (on the y axis).  The observations for each superorbital phase bin are shown in different colours as filled symbols and offset for clarity.    \label{supero2:fig:checks_1909}
  }
\end{figure}

 \tabcolsep 4pt
 \begin{table}	
 \begin{center}  
   \caption{Orbital and superorbital phase-resolved spectral analysis. \label{supero2:tab:4u_s_o_phase_specfits} } 	
  \small
    \begin{tabular}{ccccr}
 \hline
 \noalign{\smallskip}
   Phase          & $N_{\rm H}$   & $\Gamma$ & F$_{\rm 0.3-10\,keV}$                            &  Cstat                        \cr
     range                  & $10^{22}$   &                     & $10^{-11}$ &     /d.o.f.      \cr 
                                 &  (cm$^{-2}$)  &                     & (erg\,cm$^{-2}$\,s$^{-1}$) &     \\   
\noalign{\smallskip} 
\hline   
\noalign{\smallskip}
 Orbital & & & & \\ 
 \hline
 \noalign{\smallskip}
0.000--0.083 & $13.4^{+1.5}_{-1.4}$ & $1.2^{+0.2}_{-0.2}$  & $15.2^{+0.8}_{-0.8}$   &     610.9/640\\
0.083--0.167 & $10.9^{+1.7}_{-1.5}$ & $1.0^{+0.3}_{-0.2}$  & $17.1^{+1.2}_{-1.1}$   &     546.1/578\\
0.167--0.250 & $9.9^{+1.2}_{-1.1}$ & $1.2^{+0.2}_{-0.2}$  & $13.0^{+0.7}_{-0.7}$   &     633.9/659\\
0.250--0.333  & $9.0^{+0.9}_{-0.9}$ & $1.2^{+0.2}_{-0.2}$  & $17.2^{+0.8}_{-0.8}$   &     659.7/704\\
0.333--0.417 & $8.5^{+1.0}_{-1.0}$ & $0.9^{+0.2}_{-0.2}$  & $21.7^{+1.1}_{-1.1}$   &     640.1/693\\
0.417--0.500 & $8.3^{+1.0}_{-0.9}$ & $1.0^{+0.2}_{-0.2}$  & $23.3^{+1.2}_{-1.1}$   &     698.9/682\\
0.500--0.583 & $10.4^{+1.2}_{-1.1}$ & $1.0^{+0.2}_{-0.2}$  & $15.4^{+0.7}_{-0.7}$   &     668.1/718\\
0.583--0.667 & $11.3^{+1.5}_{-1.4}$ & $0.8^{+0.2}_{-0.2}$  & $12.7^{+0.7}_{-0.6}$   &     597.2/676\\
0.667--0.750 & $15.7^{+2.7}_{-2.4}$ & $1.1^{+0.3}_{-0.3}$  & $13.7^{+1.0}_{-0.9}$   &     475.6/559\\
0.750--0.833 & $27.7^{+5.8}_{-5.2}$ & $1.2^{+0.5}_{-0.4}$  & $7.5^{+0.7}_{-0.6}$   &     439.8/480\\
0.833--0.917 & $18.9^{+4.1}_{-3.6}$ & $0.4^{+0.3}_{-0.3}$  & $8.8^{+0.7}_{-0.6}$   &     544.6/557\\
0.917--0.999  & $19.5^{+2.5}_{-2.3}$ & $1.1^{+0.3}_{-0.2}$  & $10.8^{+0.6}_{-0.6}$   &     579.8/649\\ 
\noalign{\smallskip} 

      \noalign{\smallskip} 
\hline   
\noalign{\smallskip} 
Superorbital & & & & \\ 
 \hline
 \noalign{\smallskip}
0.000--0.083 & $12.8^{+2.1}_{-1.9}$ & $0.8^{+0.3}_{-0.3}$  & $9.4^{+0.7}_{-0.6}$   &     532.1/577\\
0.083--0.167 & $9.1^{+1.3}_{-1.2}$ & $1.0^{+0.2}_{-0.2}$  & $13.3^{+0.9}_{-0.8}$   &     520.0/609\\
0.167--0.250 & $12.1^{+1.8}_{-1.6}$ & $1.0^{+0.2}_{-0.2}$  & $10.6^{+0.7}_{-0.6}$   &     599.1/609\\
0.250--0.333 & $9.5^{+1.2}_{-1.1}$ & $0.7^{+0.2}_{-0.2}$  & $18.3^{+1.0}_{-0.9}$   &     695.2/666\\
0.333--0.417 & $9.9^{+1.2}_{-1.1}$ & $0.8^{+0.2}_{-0.2}$  & $15.8^{+0.8}_{-0.7}$   &     708.6/701\\
0.417--0.500 & $11.2^{+1.6}_{-1.5}$ & $0.8^{+0.2}_{-0.2}$  & $21.2^{+1.3}_{-1.2}$   &     562.9/636\\
0.500--0.583 & $9.7^{+1.5}_{-1.3}$ & $0.7^{+0.2}_{-0.2}$  & $14.7^{+0.9}_{-0.8}$   &     631.8/645\\
0.583--0.667 & $12.3^{+1.4}_{-1.3}$ & $1.2^{+0.2}_{-0.2}$  & $11.9^{+0.6}_{-0.6}$   &     635.5/672\\
0.667--0.750 & $10.1^{+1.7}_{-1.5}$ & $1.1^{+0.3}_{-0.3}$  & $12.6^{+1.0}_{-0.9}$   &     462.6/540\\
0.750--0.833 & $9.6^{+1.2}_{-1.1}$ & $0.9^{+0.2}_{-0.2}$  & $19.4^{+1.0}_{-1.0}$   &     691.7/675\\
0.833--0.917 & $10.4^{+1.4}_{-1.3}$ & $0.9^{+0.2}_{-0.2}$  & $15.0^{+0.8}_{-0.8}$   &     701.9/660\\
0.917--0.999 & $9.8^{+1.2}_{-1.1}$ & $0.8^{+0.2}_{-0.2}$  & $14.9^{+0.7}_{-0.7}$   &     674.7/709\\
\noalign{\smallskip}
  \hline
  \end{tabular}

\end{center}
\end{table}

The XRT data were processed and analysed uniformly by using 
{\sc FTOOLS}\footnote{\href{https://heasarc.gsfc.nasa.gov/ftools/ftools_menu.html}{https://heasarc.gsfc.nasa.gov/ftools/ftools\_menu.html.}} (v6.34), and matching calibration (CALDB\footnote{\href{https://heasarc.gsfc.nasa.gov/docs/heasarc/caldb/caldb_intro.html}{https://heasarc.gsfc.nasa.gov/docs/heasarc/caldb/caldb\_intro.html.}}) files. We used {\sc XSPEC} \citep[][v12.14.1]{Arnaud1996:xspec} for the spectral analysis and adopted  C-statistics \citep[][]{Cash1979}. We assumed an absorbed power-law model with free absorption and photon index as model.  The absorption component was modelled with {\sc tbabs}  with the default \citet[]{Wilms2000} abundances ({\sc abund wilm}) and \citet[][]{Verner1996:xsecs} cross sections ({\sc xsect vern}). Based on the known properties of the source spectral energy distribution in the 0.5--10~keV energy domain (see Sect.~\ref{supero2:4u}), such a model proved to be able to satisfactorily describe the XRT spectra, with no significant discrepancy identified in the residuals from the different fits that could suggest the need for an additional spectral component and/or the adoption of a different phenomenological spectral model. In this paper, errors on the spectral parameters are reported at the 90\,\% confidence level (c.l.). The procedure we adopted for the \sw/XRT data analysis is detailed in \citet[][]{Romano2024:superorbital}, and summarised here. 

We first analysed each individual observation. We calculated the orbital (superorbital) phase and from the count rates in the 0.3--4 and 4--10\,keV energy bands, we then calculated the HR = CR(4--10)/CR(0.3--4) hardness ratio. Subsequently, we extracted the average spectrum and fitted it with an absorbed power law to measure the observed 0.3--10\,keV flux (Col.\ 8 of Table~\ref{supero2:tab:swift_xrt_log_u1909}).

\begin{figure*}

 \hspace{-0.5truecm}
   \includegraphics[width=18.5cm,angle=0]{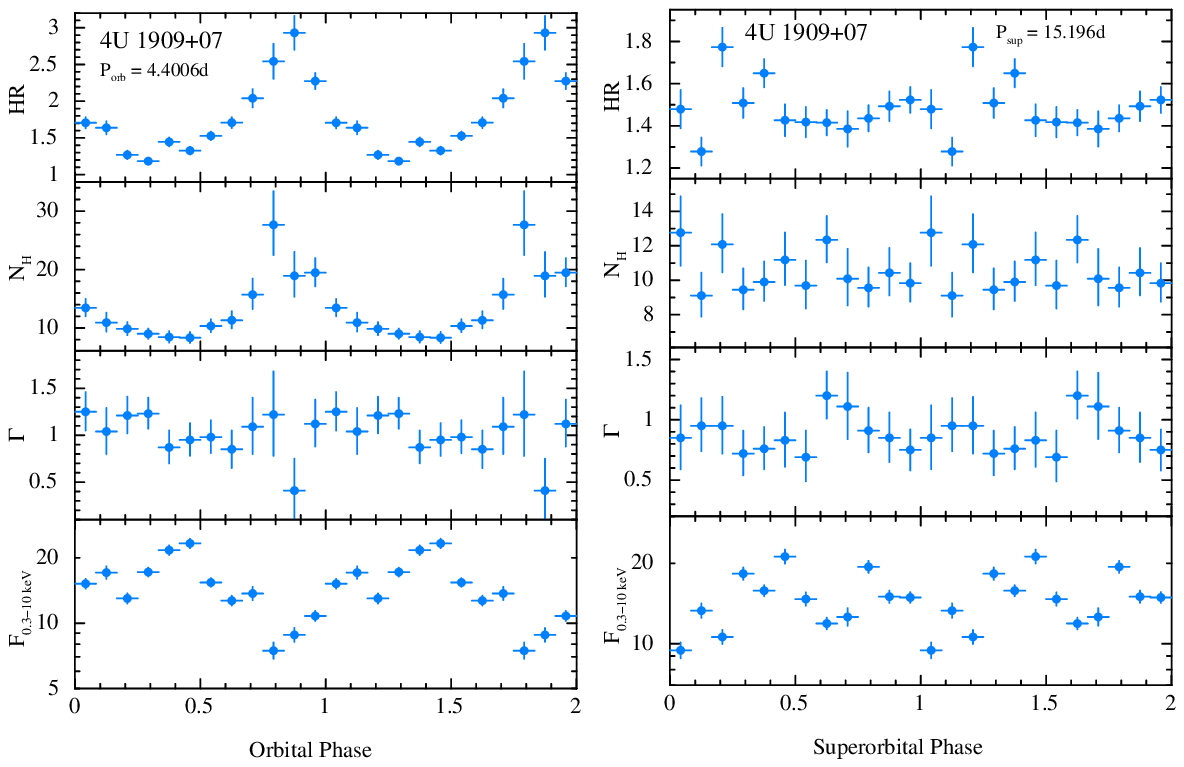}
  
  \vspace{-14.5truecm}
  \caption{ 
  {\it Left}: \sw/XRT hardness ratio of \mysrc\ and best-fit parameters as a function of orbital phase (P$_{\rm orb}=4.4006\pm0.0006$\,d, $T_{0}$= MJD\,50439.919). The absorption column density N$_{\rm H}$ is in units of 10$^{22}$ cm$^{-2}$, the power-law photon index is $\Gamma$, and the observed 0.3--10 keV flux is  in units of 10$^{-11}$ erg cm$^{-2}$ s$^{-1}$). {\it Right}: same as left side but for the superorbital phase (P$_{\rm sup}$=15.196$\pm$0.004\,d, $T_{0}$= MJD\,59502.1).   \label{supero2:fig:4u1909_orbital_phase_fits_campaign}
  }
\end{figure*}

Second, we performed a phase-resolved spectroscopy on the whole campaign by using 12 orbital (superorbital) phase bins, 12 being a compromise between achieving a sufficiently high signal-to-noise ratio spectrum to accurately determine both the absorption column density and the power-law slope and having enough points to follow possible variations along the cycles \citep[see also the discussion in our previous papers,][]{Ferrigno2022:SGXRB,Romano2024:superorbital}. As a result, our bins average $\approx 2400$ cts bin$^{-1}$. We calculated the HR in those bins, as shown in Fig.~\ref{supero2:fig:4u1909_orbital_phase_fits_campaign} top. We then created an average spectrum in each phase bin and, by adopting an absorbed power-law model, we fit it in the 0.3--10\,keV energy range. Our results are summarised in Table~\ref{supero2:tab:4u_s_o_phase_specfits} and plotted in Fig.~\ref{supero2:fig:4u1909_orbital_phase_fits_campaign} (left for the orbital, right for the superorbital case). 

We note that, since the source pulse period is $\sim605$\,s,  shorter than the typical XRT exposure, any spectral energy variability due to the energy-dependent pulsations from the source is averaged out in the stacked XRT spectra and does not affect the results of the orbital and superorbital spectral variability investigations we discuss. The validity of this procedure was extensively discussed already in our previous paper focusing on the wind-fed X-ray binary 2S~0114+650 \citep{Romano2024:superorbital}, and was also and verified with the XRT on-line analysis tool \citep{Evans2009:xrtgrb_mn}.

       \subsection{\emph{Swift}/BAT}  \label{supero2:data_bat} 

        The \sw/BAT Transient Monitor\footnote{\href{http://swift.gsfc.nasa.gov/docs/swift/results/transients/}{http://swift.gsfc.nasa.gov/docs/swift/results/transients/. }} \citep[][]{Krimm2013:BATTM} has been providing monitoring of more than a thousand sources, including high-mass X-ray binaries, in the 15--50\,keV energy band  for over 20 years.  To investigate the behaviour and potential modulation of the orbital and super-orbital periods of \mysrc,  we used light-curves from the BAT Transient Monitor\footnote{\href{https://swift.gsfc.nasa.gov/results/transients/weak/4U1909p07/}{https://swift.gsfc.nasa.gov/results/transients/weak/4U1909p07/.}} binned at the {\it Swift}-orbit level from MJD 53416 to MJD 60670 (thus including the time period of the XRT observational campaign of \mysrc). The light-curves were further screened to exclude bad quality points and to only use the data for which the data-quality flag `DATA\_FLAG' was set to 0. A small number of data points with very low fluxes and unrealistically small uncertainties were also removed from the light-curves.  The times were corrected to the solar sy We constructed a dynamic power spectrum, exploiting the technique illustrated in \citet{Islam2023_superorbital}. The power spectrum was calculated using 750 day intervals that were successively shifted in time by 50 days relative to each other, using the semi-weighting technique, in which the error bar on each data point and the excess variability in the light-curve were taken into account \citep{CorbetKrimm2013:superorbital,Corbet2007}. 
        
\begin{figure*}
            \centering
           \includegraphics[width=1.0\linewidth]{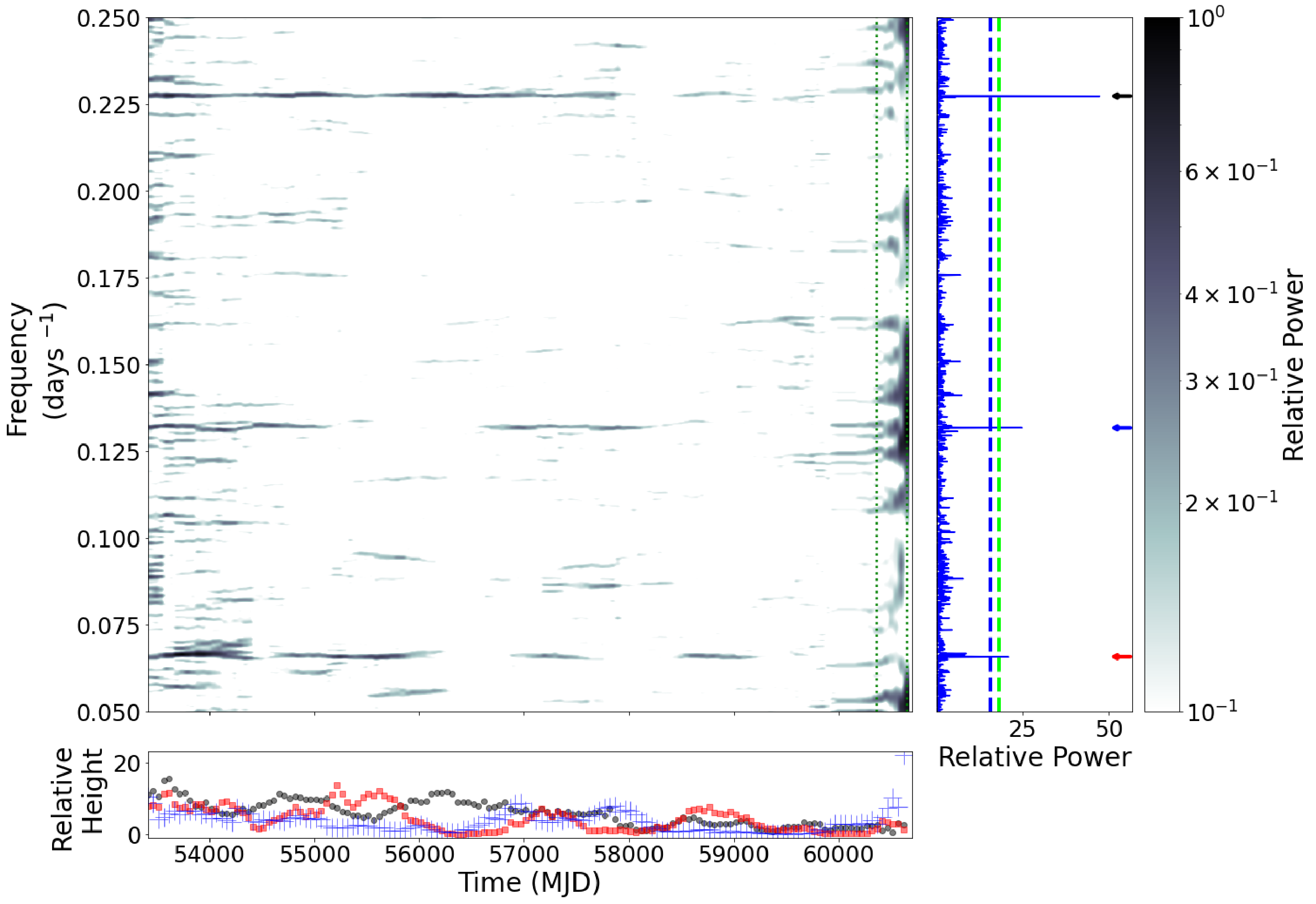}
            \caption{\sw/BAT (15--50\,keV) dynamic power spectrum (DPS) analysis of \mysrc\ of the entire observation period (MJD\,53416--60670). 
              {\it Left panel:} DPS from 0.05 to 0.25\,d$^{-1}$.  The vertical dotted lines show the time range of the XRT campaign reported in this paper.
              {\it Right panel:} orbital period (4.4\,d, indicated by black arrow) and superorbital period of 15.196\,d ($\sim0.06$\,d$^{-1}$, red arrow) and its second harmonic (blue arrow).   
            The white noise 99.9\,\% and 99.99\,\% significance levels are indicated by the dashed blue and green lines, respectively. {\it Lower panel:} relative strengths of the orbital period (black), superorbital period (red) and its second harmonic (blue). }
            \label{fig:dps_bat1}
         \end{figure*}

Our results are shown in top panel of Fig.~\ref{fig:dps_bat1} which exhibits peaks at the orbital period of 4.4\,d, superorbital periods at the fundamental frequency of 15.196\,d and its second harmonic of 7.598\,d.  The lower panel of Fig.~\ref{fig:dps_bat1} shows the relative strengths of the orbital modulation, the fundamental and the second harmonic of the superorbital modulations while the right panel shows the power spectrum and the different peaks corresponding to orbital and the fundamental frequency and the second harmonic of the superorbital modulations. As seen in Fig.~\ref{fig:dps_bat1}, both the fundamental frequency and the second harmonic of the superorbital modulations are present from MJD 53416 to MJD 55500, whereas either the fundamental period of the superorbital modulations or the second harmonic is present from MJD 55500 to MJD 59000, similarly to that reported in \cite{Islam2023_superorbital}. From MJD 59500, both the superorbital modulations at the fundamental period and the second harmonic became weaker and are consistent with the mean power. The vertical dashed lines in the plot show the time period of the XRT campaign reported in this paper. The superorbital modulations both at the fundamental period and the second harmonic became weaker during the XRT campaign. After MJD 60000, near the end of the dynamic power spectrum, there are several spurious peaks in the dynamic power spectrum which have a low statistical significance. This is likely due to aliasing effect arising at the end of the last sliding window used to construct the dynamic power spectrum.        

   \section{Results and discussion} \label{supero2:discussion}  
 
In this paper, we report on a ten-month observational campaign performed on  \mysrc\ with \sw/XRT to investigate the spectral variability along its orbital and superorbital periodicities. A visual summary of the outcomes of the analysis of the XRT data is shown in Fig.~\ref{supero2:fig:4u1909_orbital_phase_fits_campaign} (left for orbital, right for superorbital). By comparing the left side of this figure with Fig.~6 and 8 from \citet{Levine2004:1909}, we can see that the XRT observations were effective in revealing a similar variability in the source power-law slope and absorption column density along the orbit that was already identified in the past with the \rxte/PCA (although this instrument, contrary to XRT, did not provide data coverage at energies 0.5--2 keV). 

The modulated flux shows a dip around phase 1, a power-law slope that remains virtually constant across all phases, and an absorption column density which substantially increases in correspondence to the flux dip. This dip could be satisfactorily explained by \citet{Levine2004:1909} as being due to the passage of the neutron star behind the companion, providing constraints on the binary parameters (masses of the primary and secondary star, as well as inclination, eccentricity, mass loss rate, and semi-major axis) based on the evidence of the lack of a full eclipse. The XRT data allowed us to constrain the absorption column density at the level of few \% as a function of the orbital phase and confirm the lack of any significant variability in the slope of the continuum emission (to within the recorded uncertainties of $\pm$20--30\% on the $\Gamma$ parameter). 

The right side of Fig.~\ref{supero2:fig:4u1909_orbital_phase_fits_campaign} shows, for the first time, the soft X-ray light-curve of \mysrc\ folded on the superorbital period. Contrary to the orbital plot on the left, there seems to be no significant variability in the spectral properties of the source at the different superorbital phases. The XRT data show that both the power-law photon index and the absorption column density remain virtually constant across all phases, with only the X-ray flux displaying noticeable variations. These appear to be rather stochastic,  with no obviously recognizable pattern that could suggest a specific dependence of the flux from the superorbital phases. It is interesting to compare this figure with Fig.~3 of \citet{Islam2023_superorbital}, i.e.\ the variability of the source luminosity at different superorbital phases measured with \sw/BAT in the hard X-ray domain (15--50\,keV; note that we used the same value of the source orbital period and reference epoch to fold the XRT light-curve to ease the comparison). The folded BAT light-curve displayed a smoother variability with an apparent (but barely significant) sinusoidal-like modulation only during the time intervals when the fundamental frequency of the superorbital modulation is far stronger than the second harmonic and both of them are clearly detected in the dynamic power spectrum (e.g., during the time interval MJD\,53416--55500). During the time intervals when the second harmonic is detected with a higher power than the fundamental, or when both frequencies are either marginally or not detected at all, then the fractional root mean square (RMS) amplitude of the superorbital intensity profiles increases and the folded BAT light-curve has a more stochastic behaviour. Figure~\ref{fig:dps_bat1}  shows that the BAT data accumulated during the period corresponding to the XRT campaign (MJD\,60358--60649) are characterised by 
aliasing effects due to the end of the last sliding window and increasing background noise in the recent BAT data and neither the fundamental nor the second harmonic of the superorbital modulation are clearly detected. In this context, it is likely that the stochastic behavior of the folded XRT light-curve can be ascribed to the lack of a sufficiently strong superorbital modulation and this can explain the lack of any significant phase-dependent variation in the source absorption column density and photon index (see Fig.~\ref{supero2:fig:4u1909_orbital_phase_fits_campaign}, right). 

The strong suppression of the superorbital modulation across different time intervals of a wind-fed binary is still a phenomenon that remains largely unexplored and not understood \citep[see discussion in][]{Islam2023_superorbital}. In the context of the CIR model, it is possible to ascribe the strong suppression of the modulation to a substantial change in the properties of the CIRs surrounding the massive star, which might change or completely dissolve over time (to be, eventually, reformed at a later stage; see also \citealt[][for a discussion about the possible relevance of the tidal oscillation model]{Koenigsberger2006,Coley2019:16493superorbital,Islam2023_superorbital}). To the best of our knowledge, it cannot be completely ruled out that a CIR can evolve on time scales of years and thus affect, according to the CIR model for the superorbital modulation, the intensity variations displayed by a wind-fed binary. As discussed by \citet{Bozzo2017:superorbital}, the formation and evolution over long time scales ($\sim$years) of the CIRs is a field yet to be substantially developed due to the lack of observational data in the UV domain following the decommissioning of the {\it International Ultraviolet Explorer} \citep[IUE,][]{Faelker1987:IUE_procs} satellite in 1996. Data from the IUE satellite have been so far the primary (and only) source of discovery and characterization of the so-called discrete absorption features (DACs) in the spectra of supergiant stars which led to the formulation of the CIR theory \citep[see, e.g.,][for a recent review]{Driessen2023:dacs}. 

A further aspect to be considered is that CIRs might not need to be completely dissolved into the surrounding wind for the superorbital modulation of the X-ray emission from supergiant wind-fed X-ray binary to become undetectable. More modest changes in the CIRs velocity and density contrast against the surrounding stellar wind might indeed have just decreased to below the level required for us to be sensitive to measurements of the induced variability on the intensity and spectral energy distribution. If the CIRs are only mildly over-dense and moderately slower/faster than the surrounding wind, then the increase in the mass accretion rate occurring when the compact object is crossing one of these structures, together with possible changes in the local absorption column density, could only be limited in amplitude and well hidden within the uncertainties of the XRT measurements (up to 20--30\,\% on both the $N_{\rm H}$ and $\Gamma$ parameters, see Fig.~\ref{supero2:fig:4u1909_orbital_phase_fits_campaign}).

\begin{acknowledgements}
We acknowledge unwavering support from Heidi. 
This research was funded by the "Programma di Ricerca Fondamentale INAF 2023". 
This work has been partially supported by the ASI-INAF program I/004/11/5.  
NI acknowledges the NASA grant 80NSSC25K7622. 
This work was supported in part by NASA under award number 80GSFC24M0006. 
The data underlying this article are publicly available from the \sw\ archive.  
This work made use of data supplied by the UK Swift Science Data Centre at the University of Leicester.
Happy 20th, \sw. 
\end{acknowledgements}

\bibliographystyle{aa}


\begin{appendix}
  \FloatBarrier 
\twocolumn
 
\setcounter{table}{0}		
  \begin{table*}[h!]
\section{Supplementary tables}
\begin{center} 
  \caption{Observation log for 4U~1909$+$07: ObsID, date (MJD of the middle of the observation), orbital phase, superorbital phase, start and end times (UTC), XRT exposure time, 
and  0.3--10\,keV observed  flux. \label{supero2:tab:swift_xrt_log_u1909} } 	
  \scriptsize 
    \begin{tabular}{cc ll ll cc}
 \hline
 \noalign{\smallskip}
      ObsID   & MJD           &  Orbital$^a$   &  Super-o.$^b$   &    Start time (UT)      &  End time (UT)    &  Exposure & Flux$^c$ \\
        &     & Phase          &    Phase           &   (yyyy-mm-dd hh:mm:ss) & (yyyy-mm-dd hh:mm:ss)   &  (s)   & (erg\,cm$^{-2}$\,s$^{-1}$) \\
  \noalign{\smallskip}
 \hline
 \noalign{\smallskip}
00016436001 & 60358.71875	& 0.97	 &      0.37 & 2024-02-18  17:05:00  & 2024-02-18  17:19:52 & 	893	& $15.6^{+2.9}_{-2.3}$ \\ 
00016436002 & 60360.45703	& 0.36	 &      0.49 & 2024-02-20  05:38:00  & 2024-02-20  16:14:53 & 	634	& $2.9^{+1.8}_{-1.1}$ \\ 
00016436003 & 60362.93359	& 0.92	 &      0.65 & 2024-02-22  22:19:16  & 2024-02-22  22:32:54 & 	817	& $10.3^{+3.0}_{-2.3}$ \\ 
00016436004 & 60365.50000	& 0.51	 &      0.82 & 2024-02-25  11:51:22  & 2024-02-25  12:05:52 & 	870	& $18.7^{+3.7}_{-3.1}$ \\ 
00016436006 & 60370.57812	& 0.66	 &      0.15 & 2024-03-01  13:44:02  & 2024-03-01  13:58:52 & 	890	& $15.8^{+3.6}_{-3.0}$ \\ 
00016436007 & 60372.34375	& 0.06	 &      0.27 & 2024-03-03  08:09:20  & 2024-03-03  08:18:52 & 	572	& $28.4^{+5.1}_{-4.2}$ \\ 
00016436008 & 60375.98047	& 0.89	 &      0.51 & 2024-03-06  23:22:40  & 2024-03-06  23:38:53 & 	973	& $6.6^{+1.8}_{-1.4}$ \\ 
00016436009 & 60377.88281	& 0.32	 &      0.63 & 2024-03-08  21:01:39  & 2024-03-08  21:20:32 & 	1133	& $19.2^{+2.8}_{-2.4}$ \\ 
00016436010 & 60379.40234	& 0.67	 &      0.73 & 2024-03-10  09:33:18  & 2024-03-10  09:48:54 & 	935	& $13.2^{+2.8}_{-2.2}$ \\ 
00016436011 & 60382.03516	& 0.26	 &      0.91 & 2024-03-13  00:42:21  & 2024-03-13  00:56:54 & 	873	& $13.8^{+3.1}_{-2.4}$ \\ 
00016436012 & 60384.32812	& 0.79	 &      0.06 & 2024-03-15  07:53:12  & 2024-03-15  07:55:52 & 	160	& $9.2^{+13.2}_{-4.9}$ \\ 
00016436020 & 60404.62891	& 0.40	 &      0.39 & 2024-04-04  14:57:23  & 2024-04-04  15:11:53 & 	870	& $35.7^{+6.2}_{-5.2}$ \\ 
00016436021 & 60406.59766	& 0.85	 &      0.52 & 2024-04-06  14:11:44  & 2024-04-06  14:25:52 & 	847	& $4.6^{+1.5}_{-1.1}$ \\ 
00016436022 & 60409.83203	& 0.58	 &      0.73 & 2024-04-09  19:50:45  & 2024-04-09  20:05:53 & 	908	& $13.1^{+2.4}_{-2.0}$ \\ 
00016436023 & 60411.54297	& 0.97	 &      0.85 & 2024-04-11  06:41:47  & 2024-04-11  19:22:52 & 	1417	& $18.1^{+2.9}_{-2.5}$ \\ 
00016436024 & 60414.60156	& 0.66	 &      0.05 & 2024-04-14  10:23:36  & 2024-04-14  18:31:52 & 	1118	& $15.5^{+3.4}_{-2.8}$ \\ 
00016436025 & 60416.15625	& 0.02	 &      0.15 & 2024-04-16  03:40:26  & 2024-04-16  03:54:54 & 	868	& $14.3^{+3.1}_{-2.5}$ \\ 
00016436028 & 60423.32422	& 0.65	 &      0.62 & 2024-04-23  07:40:11  & 2024-04-23  07:53:54 & 	822	& $4.3^{+1.2}_{-0.9}$ \\ 
00016436029 & 60427.17969	& 0.52	 &      0.88 & 2024-04-26  16:17:13  & 2024-04-27  16:19:52 & 	1740	& $15.2^{+2.1}_{-1.8}$ \\ 
00016436030 & 60428.59375	& 0.84	 &      0.97 & 2024-04-28  01:27:52  & 2024-04-29  03:00:52 & 	2219	& $13.1^{+1.7}_{-1.5}$ \\ 
00016436031 & 60431.09375	& 0.41	 &      0.13 & 2024-05-01  02:09:54  & 2024-05-01  02:24:52 & 	898	& $12.2^{+2.2}_{-1.9}$ \\ 
00016436032 & 60433.73438	& 0.01	 &      0.31 & 2024-05-03  17:28:50  & 2024-05-03  17:42:53 & 	842	& $20.1^{+3.5}_{-2.9}$ \\ 
00016436033 & 60436.22656	& 0.58	 &      0.47 & 2024-05-06  05:21:11  & 2024-05-06  05:34:53 & 	822	& $29.9^{+5.8}_{-4.8}$ \\ 
00016436034 & 60438.91797	& 0.19	 &      0.65 & 2024-05-08  21:52:32  & 2024-05-08  22:05:52 & 	800	& $2.4^{+0.7}_{-0.5}$ \\ 
00016436035 & 60440.30859	& 0.51	 &      0.74 & 2024-05-10  04:10:55  & 2024-05-10  10:33:52 & 	855	& $4.4^{+1.4}_{-1.0}$ \\ 
00016436036 & 60443.45703	& 0.22	 &      0.95 & 2024-05-13  10:50:49  & 2024-05-13  11:05:52 & 	903	& $17.6^{+2.8}_{-2.4}$ \\ 
00016436038 & 60445.34766	& 0.65	 &      0.07 & 2024-05-15  04:23:41  & 2024-05-15  12:16:53 & 	780	& $9.5^{+2.2}_{-1.8}$ \\ 
00016436040 & 60448.92969	& 0.47	 &      0.31 & 2024-05-18  22:11:27  & 2024-05-18  22:26:52 & 	925	& $15.4^{+4.4}_{-3.3}$ \\ 
00016436041 & 60450.36719	& 0.79	 &      0.40 & 2024-05-20  08:43:09  & 2024-05-20  08:55:53 & 	765	& $4.4^{+1.5}_{-1.2}$ \\ 
00016436043 & 60453.56641	& 0.52	 &      0.61 & 2024-05-23  05:05:34  & 2024-05-23  22:06:53 & 	973	& $13.0^{+2.3}_{-1.9}$ \\ 
00016436044 & 60455.96484	& 0.06	 &      0.77 & 2024-05-25  23:02:24  & 2024-05-25  23:16:52 & 	868	& $20.6^{+4.5}_{-3.5}$ \\ 
00097732002 & 60457.66797	& 0.45	 &      0.88 & 2024-05-27  15:54:25  & 2024-05-27  16:03:52 & 	567	& $14.4^{+4.0}_{-3.0}$ \\ 
00016436045 & 60458.26953	& 0.59	 &      0.92 & 2024-05-28  04:56:50  & 2024-05-28  07:54:53 & 	1058	& $18.9^{+3.2}_{-2.7}$ \\ 
00097732003 & 60460.75781	& 0.15	 &      0.09 & 2024-05-30  18:06:46  & 2024-05-30  18:20:53 & 	847	& $26.7^{+4.6}_{-3.9}$ \\ 
00016436046 & 60460.96484	& 0.20	 &      0.10 & 2024-05-30  22:59:49  & 2024-05-30  23:13:52 & 	842	& $19.2^{+4.1}_{-3.3}$ \\ 
00016436047 & 60462.74609	& 0.60	 &      0.22 & 2024-06-01  17:48:34  & 2024-06-01  18:03:51 & 	918	& $5.5^{+1.5}_{-1.2}$ \\ 
00097732004 & 60464.76953	& 0.06	 &      0.35 & 2024-06-03  18:23:15  & 2024-06-03  18:37:53 & 	878	& $8.8^{+1.6}_{-1.4}$ \\ 
00016436048 & 60465.90625	& 0.32	 &      0.42 & 2024-06-04  21:36:15  & 2024-06-04  21:49:52 & 	817	& $17.0^{+4.2}_{-3.2}$ \\ 
00097732005 & 60467.67969	& 0.73	 &      0.54 & 2024-06-06  16:11:57  & 2024-06-06  16:26:52 & 	895	& $8.7^{+2.5}_{-1.9}$ \\ 
00016436049 & 60470.95703	& 0.47	 &      0.76 & 2024-06-09  22:50:45  & 2024-06-09  23:03:53 & 	787	& $19.0^{+5.4}_{-4.2}$ \\ 
00097732006 & 60471.74219	& 0.65	 &      0.81 & 2024-06-10  17:43:14  & 2024-06-10  17:55:53 & 	760	& $5.0^{+1.2}_{-1.0}$ \\ 
00016436050 & 60472.67188	& 0.86	 &      0.87 & 2024-06-11  09:39:52  & 2024-06-11  22:37:52 & 	697	& $8.7^{+4.7}_{-2.9}$ \\ 
00097732007 & 60474.50391	& 0.28	 &      0.99 & 2024-06-13  11:58:42  & 2024-06-13  12:12:52 & 	850	& $25.4^{+4.3}_{-3.6}$ \\ 
00016436051 & 60475.42578	& 0.49	 &      0.05 & 2024-06-14  10:06:41  & 2024-06-14  10:18:53 & 	732	& $16.7^{+3.1}_{-2.6}$ \\ 
00016436052 & 60477.60938	& 0.98	 &      0.20 & 2024-06-16  14:31:33  & 2024-06-16  14:45:53 & 	860	& $11.5^{+3.0}_{-2.4}$ \\ 
00097732008 & 60478.72656	& 0.24	 &      0.27 & 2024-06-17  17:20:27  & 2024-06-17  17:35:52 & 	925	& $10.2^{+1.9}_{-1.6}$ \\ 
00016436053 & 60480.48828	& 0.64	 &      0.38 & 2024-06-19  11:37:35  & 2024-06-19  11:50:52 & 	797	& $19.0^{+3.9}_{-3.1}$ \\ 
00016436054 & 60482.21094	& 0.03	 &      0.50 & 2024-06-21  04:58:48  & 2024-06-21  05:13:54 & 	905	& $23.5^{+4.0}_{-3.4}$ \\ 
00016436055 & 60484.17188	& 0.47	 &      0.63 & 2024-06-23  03:58:35  & 2024-06-23  04:11:53 & 	797	& $21.9^{+4.7}_{-3.8}$ \\ 
00097732010 & 60485.08984	& 0.68	 &      0.69 & 2024-06-24  02:03:55  & 2024-06-24  02:17:52 & 	837	& $11.1^{+3.3}_{-2.3}$ \\ 
00016436056 & 60487.47266	& 0.22	 &      0.84 & 2024-06-26  11:13:22  & 2024-06-26  11:27:52 & 	870	& $17.6^{+2.9}_{-2.5}$ \\ 
00097732011 & 60488.65234	& 0.49	 &      0.92 & 2024-06-27  15:32:57  & 2024-06-27  15:49:52 & 	1015	& $25.4^{+4.3}_{-3.7}$ \\ 
00016436057 & 60489.83594	& 0.76	 &      1.00 & 2024-06-28  19:53:20  & 2024-06-28  20:09:53 & 	993	& $4.4^{+1.1}_{-0.8}$ \\ 
00097732012 & 60492.69531	& 0.41	 &      0.19 & 2024-07-01  11:00:02  & 2024-07-01  22:24:53 & 	1078	& $34.1^{+4.4}_{-3.9}$ \\ 
00016436058 & 60494.43750	& 0.81	 &      0.30 & 2024-07-03  10:22:29  & 2024-07-03  10:40:47 & 	1098	& $13.5^{+3.0}_{-2.4}$ \\ 
00097732013 & 60495.67578	& 0.09	 &      0.38 & 2024-07-04  16:05:05  & 2024-07-04  16:21:53 & 	1008	& $22.0^{+3.5}_{-3.0}$ \\ 
00016436060 & 60499.55469	& 0.97	 &      0.64 & 2024-07-08  13:12:14  & 2024-07-08  13:24:52 & 	757	& $3.0^{+0.9}_{-0.7}$ \\ 
00097732014 & 60499.68359	& 1.00	 &      0.65 & 2024-07-08  16:19:57  & 2024-07-08  16:33:52 & 	835	& $14.3^{+3.5}_{-2.8}$ \\ 
00016436061 & 60501.79688	& 0.48	 &      0.79 & 2024-07-10  19:02:20  & 2024-07-10  19:15:53 & 	812	& $41.3^{+5.9}_{-5.1}$ \\ 
00097732015 & 60502.65625	& 0.67	 &      0.84 & 2024-07-11  15:38:05  & 2024-07-11  15:54:53 & 	1008	& $6.7^{+1.5}_{-1.2}$ \\ 
00016436062 & 60504.82031	& 0.17	 &      0.99 & 2024-07-13  19:32:44  & 2024-07-13  19:48:52 & 	968	& $17.4^{+2.9}_{-2.5}$ \\ 
00016436063 & 60506.53125	& 0.55	 &      0.10 & 2024-07-15  12:38:18  & 2024-07-15  12:54:53 & 	995	& $9.9^{+2.3}_{-1.8}$ \\ 
00097732016 & 60506.65625	& 0.58	 &      0.11 & 2024-07-15  15:42:01  & 2024-07-15  15:44:32 & 	150	& $18.0^{+8.2}_{-5.4}$ \\ 
00097732017 & 60509.54688	& 0.24	 &      0.30 & 2024-07-18  12:59:26  & 2024-07-18  13:11:34 & 	592	& $6.5^{+1.7}_{-1.3}$ \\ 
00016436064 & 60511.19531	& 0.61	 &      0.41 & 2024-07-20  04:34:46  & 2024-07-20  04:47:50 & 	782	& $12.9^{+3.0}_{-2.4}$ \\ 

\noalign{\smallskip}
\noalign{\smallskip}
  \hline
  \end{tabular}

Notes: $^a$ Phase zero for the orbital period is 
MJD 50439.919 \citep[][]{Wen2000:1909,Levine2004:1909}.
$^b$ Phase zero for the superorbital period is the maximum of the folded light curve at MJD 59502.1 \citep[][]{Islam2023_superorbital}. 
$^c$ Observed flux in the 0.3--10\,keV energy band in units of $\times 10^{-11}$ erg\,cm$^{-2}$\,s$^{-1}$. 
When no value is provided, the observation yielded insufficient counts to perform spectral analysis. 
\end{center}
\end{table*}

\setcounter{table}{0}		
 \begin{table*} 	
 \begin{center} 
  \caption{Continued.}	
  \scriptsize 
    \begin{tabular}{cc ll ll cc}
 \hline
 \noalign{\smallskip}
      ObsID   & MJD           &  Orbital$^a$        &  Super-o.$^b$   &    Start time (UT)                 &  End time (UT)             &  Exposure & Flux$^c$  \\
                       &                   & Phase          &    Phase           &   (yyyy-mm-dd hh:mm:ss) & (yyyy-mm-dd hh:mm:ss)      &  (s)      & (erg\,cm$^{-2}$\,s$^{-1}$)  \\
  \noalign{\smallskip}
 \hline
 \noalign{\smallskip}
 00097732018 & 60513.63281	& 0.17	 &      0.57 & 2024-07-22  15:06:07  & 2024-07-22  15:20:52 & 	885	& $20.4^{+3.6}_{-3.0}$ \\ 
00016436065 & 60514.35547	& 0.33	 &      0.61 & 2024-07-23  08:24:21  & 2024-07-23  08:37:53 & 	812	& $16.6^{+3.0}_{-2.5}$ \\ 
00097732019 & 60516.91406	& 0.91	 &      0.78 & 2024-07-25  21:46:37  & 2024-07-25  22:00:52 & 	855	& $9.7^{+2.8}_{-2.1}$ \\ 
00016436066 & 60519.87500	& 0.59	 &      0.98 & 2024-07-28  20:59:57  & 2024-07-28  21:04:43 & 	286	& $33.4^{+8.9}_{-7.1}$ \\ 
00097732020 & 60520.64844	& 0.76	 &      0.03 & 2024-07-29  15:27:55  & 2024-07-29  15:41:52 & 	837	& $17.9^{+3.9}_{-3.1}$ \\
00016436067 & 60521.12109	& 0.87	 &      0.06 & 2024-07-30  02:44:45  & 2024-07-30  02:58:52 & 	845	& $6.7^{+1.9}_{-1.7}$ \\ 
00097732021 & 60523.62109	& 0.44	 &      0.22 & 2024-08-01  14:48:02  & 2024-08-01  15:01:52 & 	830	& $19.8^{+4.1}_{-3.3}$ \\ 
00097732022 & 60527.49609	& 0.32	 &      0.48 & 2024-08-05  11:49:19  & 2024-08-05  12:03:52 & 	873	& $7.8^{+2.1}_{-1.6}$ \\ 
00097732023 & 60530.18750	& 0.93	 &      0.66 & 2024-08-08  04:23:16  & 2024-08-08  04:36:53 & 	817	& $6.4^{+2.0}_{-1.4}$ \\ 
00016436070 & 60531.16406	& 0.15	 &      0.72 & 2024-08-09  03:55:36  & 2024-08-09  03:55:51 & 	15	& $13.0^{+41.1}_{-7.9}$ \\ 
00016436071 & 60533.21094	& 0.62	 &      0.85 & 2024-08-11  04:58:08  & 2024-08-11  05:05:51 & 	464	& $24.6^{+10.3}_{-7.2}$ \\ 
00097732024 & 60534.77344	& 0.97	 &      0.96 & 2024-08-12  18:27:41  & 2024-08-12  18:41:54 & 	852	& $5.3^{+1.3}_{-1.1}$ \\ 
00097732025 & 60537.40234	& 0.57	 &      0.13 & 2024-08-15  09:29:43  & 2024-08-15  09:43:53 & 	850	& $0.6^{+0.4}_{-0.2}$ \\ 
00016436072 & 60538.72656	& 0.87	 &      0.22 & 2024-08-16  17:18:38  & 2024-08-16  17:32:53 & 	855	& $1.8^{+1.2}_{-1.4}$ \\ 
00016436073 & 60541.21484	& 0.44	 &      0.38 & 2024-08-19  05:03:01  & 2024-08-19  05:20:51 & 	1071 & $35.0^{+6.6}_{-5.5}$ \\ 
00097732026 & 60541.73438	& 0.55	 &      0.41 & 2024-08-19  17:30:19  & 2024-08-19  17:44:52 & 	873	& $33.3^{+9.2}_{-8.2}$ \\ 
00016436075 & 60543.85156	& 0.04	 &      0.55 & 2024-08-21  20:18:48  & 2024-08-21  20:31:45 & 	777	& $20.5^{+5.0}_{-3.9}$ \\ 
00097732027 & 60544.37500	& 0.15	 &      0.59 & 2024-08-22  08:53:15  & 2024-08-22  09:08:53 & 	938	& $10.2^{+2.5}_{-2.0}$ \\ 
00097732028 & 60548.51172	& 0.09	 &      0.86 & 2024-08-26  12:09:36  & 2024-08-26  12:25:53 & 	978	& $12.9^{+3.3}_{-2.6}$ \\ 
00016436077 & 60550.47266	& 0.54	 &      0.99 & 2024-08-28  06:33:59  & 2024-08-28  16:03:53 & 	735	& $15.6^{+3.0}_{-2.5}$ \\ 
00097732030 & 60551.52344	& 0.78	 &      0.06 & 2024-08-29  12:28:41  & 2024-08-29  12:43:53 & 	913	& $5.4^{+1.4}_{-1.1}$ \\ 
00016436078 & 60554.08984	& 0.36	 &      0.23 & 2024-08-31  07:09:59  & 2024-09-01  21:10:52 & 	745	& $11.6^{+2.6}_{-2.1}$ \\ 
00016436079 & 60555.26562	& 0.63	 &      0.31 & 2024-09-02  06:15:49  & 2024-09-02  06:33:49 & 	1081 & $31.1^{+5.0}_{-4.3}$ \\ 
00097732031 & 60555.45703	& 0.67	 &      0.32 & 2024-09-02  10:51:08  & 2024-09-02  11:05:54 & 	885	& $30.1^{+5.7}_{-4.8}$ \\ 
00016436080 & 60558.41406	& 0.34	 &      0.51 & 2024-09-05  09:47:33  & 2024-09-05  10:01:53 & 	860	& $18.7^{+3.4}_{-2.9}$ \\ 
00097732032 & 60558.48047	& 0.36	 &      0.52 & 2024-09-05  11:22:32  & 2024-09-05  11:37:52 & 	920	& $41.6^{+6.0}_{-5.2}$ \\ 
00097732033 & 60562.41797	& 0.25	 &      0.78 & 2024-09-09  09:51:22  & 2024-09-09  10:07:52 & 	963	& $32.8^{+5.0}_{-4.3}$ \\ 
00097732034 & 60565.44531	& 0.94	 &      0.98 & 2024-09-12  10:35:10  & 2024-09-12  10:49:53 & 	883	& $14.1^{+3.0}_{-2.4}$ \\ 
00016436083 & 60565.70312	& 0.00	 &      0.99 & 2024-09-12  16:52:03  & 2024-09-12  16:56:04 & 	241	& $12.8^{+5.8}_{-4.0}$ \\ 
00016436084 & 60567.60156	& 0.43	 &      0.12 & 2024-09-14  14:23:00  & 2024-09-14  14:24:23 & 	83	& $24.1^{+15.5}_{-9.3}$ \\ 
00097732035 & 60569.38281	& 0.84	 &      0.23 & 2024-09-16  09:01:57  & 2024-09-16  09:16:52 & 	895	& $5.2^{+1.5}_{-1.2}$ \\ 
00016436085 & 60570.16016	& 0.01	 &      0.29 & 2024-09-17  03:41:44  & 2024-09-17  03:55:52 & 	847	& $13.7^{+3.8}_{-3.1}$ \\ 
00097732036 & 60572.44922	& 0.53	 &      0.44 & 2024-09-19  10:41:24  & 2024-09-19  10:54:52 & 	807	& $35.3^{+7.0}_{-5.8}$ \\ 
00016436086 & 60572.58594	& 0.57	 &      0.45 & 2024-09-19  13:57:04  & 2024-09-19  14:07:54 & 	649	& $20.4^{+4.5}_{-3.6}$ \\ 
00016436087 & 60575.60156	& 0.25	 &      0.64 & 2024-09-22  14:23:00  & 2024-09-22  14:32:52 & 	592	& $12.4^{+3.2}_{-2.5}$ \\ 
00097732037 & 60576.32031	& 0.41	 &      0.69 & 2024-09-23  07:33:05  & 2024-09-23  07:47:53 & 	888	& $18.7^{+3.6}_{-3.0}$ \\ 
00016436088 & 60577.83594	& 0.76	 &      0.79 & 2024-09-24  19:54:02  & 2024-09-24  20:08:53 & 	890	& $7.4^{+2.7}_{-1.9}$ \\ 
00097732038 & 60579.73047	& 0.19	 &      0.92 & 2024-09-26  17:23:56  & 2024-09-26  17:36:53 & 	777	& $15.5^{+3.5}_{-2.7}$ \\ 
00016436089 & 60580.19531	& 0.29	 &      0.95 & 2024-09-27  00:03:15  & 2024-09-27  09:16:54 & 	815	& $21.1^{+3.5}_{-2.9}$ \\ 
00016436090 & 60582.32031	& 0.78	 &      0.09 & 2024-09-29  06:47:20  & 2024-09-29  08:30:52 & 	624	& $8.5^{+2.6}_{-1.9}$ \\ 
00097732039 & 60583.99609	& 0.16	 &      0.20 & 2024-09-30  23:45:09  & 2024-09-30  23:59:54 & 	885	& $3.3^{+1.1}_{-0.8}$ \\ 
00016436091 & 60584.91406	& 0.37	 &      0.26 & 2024-10-01  21:50:57  & 2024-10-01  22:04:52 & 	835	& $27.8^{+4.8}_{-4.0}$ \\ 
00097732040 & 60586.29688	& 0.68	 &      0.35 & 2024-10-03  06:59:12  & 2024-10-03  07:14:53 & 	940	& $20.7^{+3.8}_{-3.2}$ \\ 
00016436092 & 60587.64062	& 0.99	 &      0.44 & 2024-10-04  14:29:21  & 2024-10-04  16:16:54 & 	920	& $18.9^{+3.6}_{-3.0}$ \\ 
00016436093 & 60589.56250	& 0.42	 &      0.56 & 2024-10-06  13:25:51  & 2024-10-06  13:37:53 & 	722	& $24.7^{+4.3}_{-3.6}$ \\ 
00097732041 & 60590.30078	& 0.59	 &      0.61 & 2024-10-07  07:06:20  & 2024-10-07  07:18:47 & 	747	& $18.9^{+5.8}_{-4.4}$ \\ 
00016436094 & 60592.51953	& 0.09	 &      0.76 & 2024-10-09  12:18:10  & 2024-10-09  12:32:52 & 	883	& $36.8^{+5.8}_{-5.0}$ \\ 
00097732042 & 60593.31641	& 0.28	 &      0.81 & 2024-10-10  07:25:14  & 2024-10-10  07:40:52 & 	938	& $13.9^{+3.2}_{-2.6}$ \\ 
00016436095 & 60594.54688	& 0.56	 &      0.89 & 2024-10-11  13:00:01  & 2024-10-11  13:11:53 & 	712	& $28.3^{+5.7}_{-4.6}$ \\ 
00016436096 & 60597.57031	& 0.24	 &      0.09 & 2024-10-14  13:40:26  & 2024-10-14  13:43:22 & 	176	& $8.0^{+4.4}_{-2.8}$ \\ 
00016436097 & 60599.26562	& 0.63	 &      0.20 & 2024-10-16  06:14:21  & 2024-10-16  06:27:54 & 	812	& $10.2^{+3.2}_{-2.4}$ \\ 
00016436098 & 60602.09375	& 0.27	 &      0.39 & 2024-10-19  02:04:52  & 2024-10-19  02:19:53 & 	898	& $27.6^{+4.5}_{-3.8}$ \\ 
00016436099 & 60604.96484	& 0.92	 &      0.58 & 2024-10-21  23:05:16  & 2024-10-21  23:18:53 & 	817	& $8.0^{+1.7}_{-1.4}$ \\ 
00016436100 & 60606.67578	& 0.31	 &      0.69 & 2024-10-23  08:22:46  & 2024-10-23  23:59:53 & 	970	& $18.9^{+3.4}_{-2.8}$ \\ 
00016436101 & 60609.09766	& 0.86	 &      0.85 & 2024-10-26  02:16:16  & 2024-10-26  02:29:56 & 	820	& $20.2^{+4.3}_{-3.4}$ \\ 
00016436103 & 60614.46484	& 0.08	 &      0.20 & 2024-10-31  11:01:10  & 2024-10-31  11:14:12 & 	424	& $9.8^{+3.5}_{-2.4}$ \\ 
00016436104 & 60616.52734	& 0.55	 &      0.34 & 2024-11-02  07:00:45  & 2024-11-02  18:23:36 & 	1201 & $14.5^{+2.3}_{-2.0}$ \\ 
00016436105 & 60619.37500	& 0.20	 &      0.52 & 2024-11-05  08:52:39  & 2024-11-05  09:05:51 & 	792	& $18.9^{+3.2}_{-2.8}$ \\ 
00016436106 & 60621.53516	& 0.69	 &      0.67 & 2024-11-07  12:44:11  & 2024-11-07  12:59:53 & 	943	& $28.7^{+4.7}_{-4.0}$ \\ 
00016436107 & 60623.76953	& 0.20	 &      0.81 & 2024-11-09  18:20:12  & 2024-11-09  18:34:52 & 	880	& $18.7^{+3.6}_{-3.0}$ \\ 
00016436108 & 60626.97266	& 0.92	 &      0.02 & 2024-11-12  23:14:53  & 2024-11-12  23:31:54 & 	1020 & $6.5^{+1.4}_{-1.1}$ \\ 
00016436109 & 60628.99609	& 0.38	 &      0.16 & 2024-11-14  23:46:08  & 2024-11-14  23:59:53 & 	825	& $25.3^{+4.3}_{-3.6}$ \\ 
00016436110 & 60631.42188	& 0.94	 &      0.32 & 2024-11-17  09:57:29  & 2024-11-17  10:11:54 & 	865	& $24.0^{+4.3}_{-3.6}$ \\ 
00016436111 & 60633.71484	& 0.46	 &      0.47 & 2024-11-19  17:00:30  & 2024-11-19  17:14:53 & 	863	& $50.2^{+7.1}_{-6.2}$ \\ 
00016436112 & 60636.45312	& 0.08	 &      0.65 & 2024-11-22  10:47:51  & 2024-11-22  10:58:52 & 	662	& $16.6^{+3.0}_{-2.5}$ \\ 
00016436113 & 60638.42969	& 0.53	 &      0.78 & 2024-11-24  10:11:03  & 2024-11-24  10:22:53 & 	710	& $32.7^{+7.0}_{-5.7}$ \\ 
00016436114 & 60641.95703	& 0.33	 &      0.01 & 2024-11-27  22:49:36  & 2024-11-27  23:03:53 & 	858	& $8.7^{+1.7}_{-1.4}$ \\ 
00016436115 & 60643.32422	& 0.64	 &      0.10 & 2024-11-29  07:41:00  & 2024-11-29  07:55:52 & 	893	& $1.3^{+1.1}_{-0.6}$ \\ 
00016436116 & 60645.02344	& 0.03	 &      0.21 & 2024-12-01  00:28:59  & 2024-12-01  00:43:52 & 	893	& $10.4^{+1.9}_{-1.6}$ \\ 
00016436117 & 60648.06641	& 0.72	 &      0.41 & 2024-12-04  00:43:49  & 2024-12-04  02:23:52 & 	830	& $8.0^{+1.7}_{-1.4}$ \\ 
\noalign{\smallskip}
\noalign{\smallskip}
  \hline
  \end{tabular}

Notes: $^a$ Phase zero for the orbital period is MJD 50439.919 \citep[][]{Wen2000:1909,Levine2004:1909}.
$^b$ Phase zero for the superorbital period is the maximum of the folded light curve at MJD 59502.1 \citep[][]{Islam2023_superorbital}. 
$^c$ Observed flux in the 0.3--10\,keV energy band in units of $\times 10^{-11}$ erg\,cm$^{-2}$\,s$^{-1}$. 
When no value is provided, the observation yielded insufficient counts to perform spectral analysis. 
\end{center}
\end{table*}

\FloatBarrier 
\clearpage

\end{appendix}
\end{document}